\documentclass[10pt]{article}
\usepackage{graphicx}

\def\Title#1{\begin{center} {\Large #1 } \end{center}}
\def\Author#1{\begin{center}{ \sc #1} \end{center}}
\def\Address#1{\begin{center}{ \it #1} \end{center}}

\newcommand\pubblock{\rightline{\begin{tabular}{l} Proceedings of the Fifth Annual LHCP\\ \pubnumber\\
         \pubdate  \end{tabular}}}

\newenvironment{Abstract}{\begin{quotation} \begin{center} 
             \large ABSTRACT \end{center}\bigskip 
      \begin{center}\begin{large}}{\end{large}\end{center} \end{quotation}}

\newenvironment{Presented}{\begin{quotation} \begin{center} 
             PRESENTED AT\end{center}\bigskip 
      \begin{center}\begin{large}}{\end{large}\end{center} \end{quotation}}

\def\Acknowledgements{\bigskip  \bigskip \begin{center} \begin{large}
             \bf ACKNOWLEDGEMENTS \end{large}\end{center}}
\newcommand{\Table}[1]{{Table~\ref{#1}}}

\newcommand{\Figure}[1]{{Figure~\ref{#1}}}

\def\tch{\textit{t}-channel}
\def\sch{\textit{s}-channel}
\def\Wt{\textit{Wt}}

\def\tZch{\textit{tZq}}

\def\mW{\ensuremath{m_{W}}}

\def\Wtb{\textit{Wtb}}
\def\flvVtb{\ensuremath{|f_{\rm LV}V_{tb}|}}
\def\Vtb{\ensuremath{|V_{tb}|}}
\def\Vts{\ensuremath{|V_{ts}|}}
\def\Vtd{\ensuremath{|V_{td}|}}

\def\ttbar{$t\bar{t}$}
\def\Met{\ensuremath{E^{\mathrm{miss}}_{\mathrm{T}}}}
\def\pt{p_{\rm T}}
\def\pT{p_{\rm T}}

\renewcommand{\Re}[1]{\mathrm{Re}\left[ #1 \right]}
\renewcommand{\Im}[1]{\mathrm{Im}\left[#1\right]}

\def\fvl{\ensuremath{f^{\mathrm{L}}_{\mathrm{V}}}}

\def\ftl{\ensuremath{f^{\mathrm{L}}_{\mathrm{T}}}}
\def\ftr{\ensuremath{f^{\mathrm{R}}_{\mathrm{T}}}}

\def\vl{\ensuremath{V_{\mathrm{L}}}}
\def\vr{\ensuremath{V_{\mathrm{R}}}}
\def\gr{\ensuremath{g_{\mathrm{R}}}}
\def\gl{\ensuremath{g_{\mathrm{L}}}}

\def\vlr{\ensuremath{V_{\mathrm{L,R}}}}
\def\glr{\ensuremath{g_{\mathrm{L,R}}}}

\def\imgr{\ensuremath{\Im{g_\mathrm{R}}}}

\newcommand{\ProjR}{\ensuremath{P_{\mathrm{R}}}}
\newcommand{\ProjL}{\ensuremath{P_{\mathrm{L}}}}

\newcommand{\sigmameas}{\ensuremath{\sigma_{\rm meas.}}}
\newcommand{\sigmatheo}{\ensuremath{\sigma_{\rm theo.}}}
\newcommand{\flv}{\ensuremath{f_{\rm LV}}}

\newcommand{\lhcenergySeven}{\ensuremath{ 7~{\rm TeV}}}
\newcommand{\lhcenergyEight}{\ensuremath{ 8~{\rm TeV}}}
\newcommand{\lhcenergyThirteen}{\ensuremath{ 13~{\rm TeV}}}
\newcommand{\cmenergySeven}{\ensuremath{\sqrt{s}=\lhcenergySeven}}
\newcommand{\cmenergyEight}{\ensuremath{\sqrt{s}=\lhcenergyEight}}
\newcommand{\cmenergyThirteen}{\ensuremath{\sqrt{s}=\lhcenergyThirteen}}
\newcommand{\cmenergyCombRunI}{\ensuremath{\sqrt{s}=7~\rm{and}~\lhcenergyEight}}
\newcommand{\cmenergyComb}{\ensuremath{\sqrt{s}=8~\rm{and}~\lhcenergyThirteen}}
\newcommand{\cmenergyCombAll}{\ensuremath{\sqrt{s}=7, 8~\rm{and}~\lhcenergyThirteen}}

\newcommand\T{\rule{0pt}{2.6ex}}       % Top strut
\newcommand\B{\rule[-1.2ex]{0pt}{0pt}} % Bottom strut
\usepackage{color}
\usepackage{lineno}
\usepackage{subfig}
\usepackage{multirow}

\usepackage{hyperref}

\newcommand\pubnumber{ ATL-PHYS-PROC-2017-124 }
\newcommand\pubdate{\today}

\def\affiliation{On behalf of the ATLAS and CMS Collaborations, \\
  Instituto de F\'{\i}sica Corpuscular (IFIC)\\
  University of Valencia and CSIC, Valencia, Spain}

\RequirePackage[normalem]{ulem} %DIF PREAMBLE
\RequirePackage{color}\definecolor{RED}{rgb}{1,0,0}\definecolor{BLUE}{rgb}{0,0,1} %DIF PREAMBLE
\begin{document}

%\linenumbers

% large size for the first page
\large
\begin{titlepage}
\pubblock

%% Change the title, name, abstract
%% Title 
\vfill
\Title{SM and BSM physics in single top quark at the LHC}
\vfill

%  if you need to add the support use this, fill the \support definition above. 
%   \Author{ FIRSTNAME LASTNAME \support }
\Author{Carlos Escobar}
\Address{\affiliation}
\vfill
\begin{Abstract}
A comprehensive review of the recent results on measurements of single top-quark
production cross-sections at \cmenergyCombAll\ performed by the ATLAS and
CMS Collaborations is presented. The cross-section measurements
include inclusive, fiducial and differential results.  In addition,
the latest measurements, based on angular distributions in \tch\ single-top-quark
processes, of the top-quark polarisation and $W$ boson spin
observables at \cmenergyEight, and the analyses of the \Wtb\ vertex at
\cmenergyCombRunI\ are also discussed. 
% These results are based on integrated luminosities of 5.1~fb$^{-1}$ at \cmenergySeven,
% 12.2 to 20.3~fb$^{-1}$ at \cmenergyEight, and 3.2~fb$^{-1}$ at
% \cmenergyThirteen.
All measurements are in good
agreement with predictions and no deviations from Standard Model
expectations have been observed so far.

\end{Abstract}
\vfill

% DO NOT CHANGE 
\begin{Presented}
The Fifth Annual Conference\\
 on Large Hadron Collider Physics \\
Shanghai Jiao Tong University, Shanghai, China\\ 
May 15-20, 2017
\end{Presented}
\vfill
\end{titlepage}
\def\thefootnote{\fnsymbol{footnote}}
\setcounter{footnote}{0}
%

% normal size for the rest
\normalsize 

%% Your paper should be entered below. 

\section{Introduction}

The top quark, which was first observed in proton--antiproton
($p\bar{p}$) collisions at the Tevatron~\cite{Aaltonen:2009jj, Abazov:2009ii}, is the heaviest known Standard Model (SM) elementary
particle. Due to its large mass~\cite{ATLAS:2014wva}, its lifetime
O($10^{-25}$~s) is smaller than its hadronisation time-scale
O($10^{-24}$~s), allowing this quark to be studied as a free
quark. Furthermore since the top-quark lifetime is also shorter than the depolarisation
timescale O($10^{-21}$~s)~\cite{Bigi:1986jk} and
the $W$ boson is produced on-shell in the top-quark decay, the
top-quark spin information is directly transferred to its decay
products. Because all of this, the top quark is fundamental for understanding
the physics in the SM and beyond. At the LHC, in
proton--proton ($pp$) collisions, top quarks are produced predominantly in
pairs (\ttbar) via the flavour-conserving strong interaction, while an
alternative process produces single top quarks through the electroweak 
interaction. Although the \ttbar\ production cross-section is larger
than that of single-top-quark production, top quarks are produced
unpolarised because of parity conservation in quantum chromodynamics
(QCD), contrary to what happens for single
top quarks. In the SM, single-top-quark production mostly proceeds via three mechanisms that can be defined at leading
order (LO): an exchange of a virtual $W$ boson either in the \tch\ or in
the \sch, or the associated production of a top quark and a $W$
boson.
%Representative Feynman diagrams of these processes, at LO in QCD.
%, are shown in \Figure{fig:feyn}.

% \begin{figure}[!htbp]
%   \centering
%   \subfloat[]{ \includegraphics[height=0.12\textheight]{tchanFeyn}}
%   \subfloat[]{ \includegraphics[height=0.12\textheight]{WtFeyn} }
%   \subfloat[]{ \includegraphics[height=0.12\textheight]{schanFeyn} }
%   \caption{Representative Feynman diagrams at LO in QCD for single-top-quark production in the (a) \tch, (b) associated \Wt\
%     production, and (c) \sch.}
%   \label{fig:feyn}
% \end{figure}

The dominant process at the LHC is the \tch, where a light-flavour quark from one
of the colliding protons interacts with a $b$-quark by exchanging a
virtual $W$ boson, producing a top quark and a recoiling light-flavour
quark q', called the spectator quark. The associated production of a
$W$ boson and a top quark has the second largest production
cross-section. This process suffers from a large background from \ttbar\
production. The \sch\ cross-section is the smallest at the LHC. The
theoretical predictions for the single-top-quark production
cross-sections are calculated at next-to-leading order
(NLO)~\cite{Aliev:2010zk, Kant:2014oha} and also at NLO evaluated
with next-to-next-to-leading logarithmic (NNLL)
resummation~\cite{Kidonakis:2010ux, Kidonakis:2012rm,
  Kidonakis:2013zqa}. A summary of the theoretical predictions of these three (total inclusive)
cross-sections, $\sigmatheo$, at \cmenergyCombAll\ is shown in
\Table{tab:predictedxsec}. The uncertainties on the theoretical predictions
include scale and PDF variations following the prescriptions
recommended by the LHC working group of top-quark physics, the
LHC$top$WG. In all cases, the top-quark mass is assumed to be
172.5 GeV; the same value which is used for the samples of simulated events.
%This channel was observed in \myppbar collisions at the
%Tevatron~\cite{CDF:2014uma} while there is only evidence for it in $pp$ collisions at the LHC at $\sqrt{s} = 8$.

% Cross-sections marked with $^\dagger$ are the ones used by the analyses presented in this review.

\begin{table}[!htbp]
  \begin{center}
    \begin{tabular}{c|c|cc}
      \multicolumn{1}{c}{} & \multicolumn{1}{c}{} & \multicolumn{2}{|c}{$\sigmatheo{}$\,(pb)} \\
      % \cline{3-4}
      \hline
      \T\B
      $\sqrt{s}$ & Process & NLO & NLO+NNLL \\
      \hline
      \T
      &\tch	                  & $63.89^{+2.91}_{-2.52}$ & $64.57^{+2.63}_{-1.74}$ \\
      \T\B
      \lhcenergySeven & \Wt & -- & $15.74^{+1.17}_{-1.21}$ \\
      \T\B
      % &\sch	                  & $4.29^{+0.19}_{-0.17}$ & $4.63^{+0.20 \dagger}_{-0.18}$ \\
      &\sch	                  & $4.29^{+0.19}_{-0.17}$ & $4.63^{+0.20}_{-0.18}$ \\
      \hline
      \T
      % &\tch	                  & $84.69^{+3.76 \dagger}_{-3.23}$ & $87.76^{+3.44}_{-1.91}$ \\
      &\tch	                  & $84.69^{+3.76}_{-3.23}$ & $87.76^{+3.44}_{-1.91}$ \\
      \T\B
      % \lhcenergyEight & \Wt & -- & $22.37\pm1.52^{\dagger}$ \\
      \lhcenergyEight & \Wt & -- & $22.37\pm1.52$ \\
      \T\B
      % &\sch	                  & $5.24^{+0.22}_{-0.20}$ & $5.61\pm0.22^{\dagger}$ \\
      &\sch	                  & $5.24^{+0.22}_{-0.20}$ & $5.61\pm0.22$ \\
      \hline
      \T
      % &\tch                                & $216.99^{+9.04 \dagger}_{-7.71}$ & -- \\
      &\tch                                & $216.99^{+9.04}_{-7.71}$ & -- \\
      \T\B
      % \lhcenergyThirteen & \Wt & -- & $71.7\pm3.8^{\dagger}$ \\
      \lhcenergyThirteen & \Wt & -- & $71.7\pm3.8$ \\
      \T\B
      &\sch	                       & $10.32^{+0.40}_{-0.36}$ & -- \\
      \hline
    \end{tabular}
    \caption{Predicted cross-sections calculated at NLO and NLO+NNLL for single-top-quark production
      at \cmenergyCombAll\ at the LHC.
      % Cross-sections marked with $^\dagger$ are the ones used by the
      % analyses presented in this review are shown.
      Uncertainties include scale and PDF variations following the prescriptions
      recommended by the LHC$top$WG.}
    \label{tab:predictedxsec}
  \end{center}
\end{table}

The production rate of single-top-quark processes is proportional to
the square of the coupling at the \Wtb\ production vertex, hence, in
the SM, the measurement of single-top-quark production cross-sections
allows for the direct determination of the magnitude of the
Cabibbo--Kobayashi--Maskawa (CKM)~\cite{Cabibbo:1963yz, Kobayashi:1973fv} matrix element, $\Vtb$. This determination does not rely on modelling assumptions
including the unitarity of the CKM matrix.

Deviations from the SM in the \Wtb\ vertex can be expressed in terms
of the (complex) anomalous couplings, $\vlr$ and $\glr$, presented by this
effective Lagrangian~\cite{AguilarSaavedra:2008zc}:
\begin{equation}
  {\cal L}_{\mathrm{eff}} = - \frac{g}{\sqrt{2}}{\overline{b}}\gamma^\mu \left( \vl \ProjL + \vr \ProjR \right) tW^-_\mu - 
  \frac{g}{\sqrt{2}}{\overline{b}}\frac{i\sigma^{\mu\nu}q_{\nu}}{\mW}
  \left( \gl \ProjL + \gr \ProjR \right) tW^-_\mu + {\rm h.c.} \,,
\end{equation}
In the SM at LO, all coupling constants vanish, except $\vl = \Vtb$. Deviations from these values would provide hints of physics
beyond the SM, and furthermore, complex values could imply that the top-quark decay has a CP-violating component~\cite{AguilarSaavedra:2008zc}.

Moreover, as a consequence of the vector--axial form of the \Wtb\ vertex in
the SM, the spin of single top quarks in \tch\ production is
predominantly aligned along the direction of the spectator-quark
momentum~\cite{Mahlon:1996pn}. Therefore, \tch\ events allow the measurement of the top-quark polarisation and all $W$ boson spin observables~\cite{Aguilar-Saavedra:2015yza}.

This review focuses on the latest results on single-top-quark analysis performed by ATLAS~\cite{Aad:2008zzm} and CMS~\cite{Chatrchyan:2008aa}
in \textit{pp} collisions at the LHC. This includes inclusive,
fiducial and differential cross-section
measurements at \cmenergyCombAll\ and measurements of the top-quark polarisation and $W$ boson spin observables, and searches of anomalous couplings in \tch\ single-top-quark processes at \cmenergyCombRunI.

\section{Single-top-quark production cross-section measurement}

The most recent \tch\ and \Wt\ single-top-quark production
cross-sections measured by ATLAS and CMS at \cmenergyComb\ are
presented in this section. Fiducial and differential
cross-section measurements are also provided for the \tch\ by ATLAS
and CMS at \cmenergyComb. Searches for the \sch\ production  at
\cmenergyCombRunI\ are shown, where the evidence of its
production is provided at \cmenergyEight\ by the ATLAS Collaboration. From these
three production cross-sections at \cmenergyCombAll,
the summary of all \Vtb\ extractions is given. Moreover, the search
of the \tZch\ production at \cmenergyThirteen\ within the SM is also presented.

\subsection{Measurement of the \tch\ production cross-section}

The event signature of the \tch\ contains a high transverse momentum ($\pT$) isolated lepton (electron or muon), missing transverse momentum
($\Met$) and two jets, where one originates from a $b$-quark
($b$-tagged jet or $b$-jet) and the other from the spectator quark. Multivariate
analysis techniques, in particular neural network (NN), are used to
separate the signal from the background, and then a binned
maximum-likelihood fit to data is performed. At \cmenergyEight, the measurement of the fiducial \tch\ production
cross-section using 19.7~fb$^{-1}$ is $3.38 \pm 0.32$~pb
in the CMS analysis~\cite{CMS:2015jca}, while the measurement
performed by the ATLAS analysis~\cite{Aaboud:2017pdi}, using
20.2~fb$^{-1}$ and a different fiducial phase space, is $9.78 \pm 0.57$~pb and $5.77 \pm 0.45$~pb, for top quark and top
antiquark respectively. The dominant systematic uncertainty is jet
energy scale (JES) in both analyses. In addition, the ratio of top-quark to
top-antiquark production cross-sections is determined to be
%R$_{t} = 1.72 \pm 0.09$,
R$_{t} = 1.72 \pm 0.05~{\rm (stat.)} \pm 0.07~{\rm (syst.)}$, 
with an improved relative precision of 4.9\% since several
systematic uncertainties cancel in the ratio. Furthermore, the ATLAS
analysis also provides the differential cross-sections as a function
of the $\pT$ and the absolute value of the rapidity ($|y|$) for both the top quark and the top
antiquark at the parton and particle levels. The
$\pt$ and $\eta$ differential cross-sections of the spectator jet from
the \tch\ scattering are also measured at particle level. The dominant
systematic uncertainties are JES and signal and \ttbar\ modelling. At \cmenergyThirteen, the early measurement of the inclusive \tch\ production
cross-section using 2.2~fb$^{-1}$ in the CMS
analysis~\cite{Sirunyan:2016cdg} is $238 \pm 13~{\rm
  (stat.)} \pm 29~{\rm (syst.)}$~pb while using 3.2~fb$^{-1}$ in the ATLAS
analysis~\cite{Aaboud:2016ymp} is $247 \pm 6~{\rm
  (stat.)} \pm 45~{\rm (syst.)}$~pb. Additionally, the ratio R$_{t}$
is $1.81 \pm 0.18~{\rm (stat.)} \pm 0.15~{\rm (syst.)}$
and $1.72 \pm 0.09~{\rm (stat.)} \pm 0.18~{\rm (syst.)}$ for the CMS and
ATLAS analyses respectively. The dominant systematic
uncertainties in the CMS analysis are the signal and \ttbar\
modelling, and the \tch\ factorisation and renormalisation scales. In
the ATLAS analysis, the dominant systematic uncertainties are the
parton shower and the $b$-tagging efficiency. Moreover, the CMS Collaboration provides the differential
cross-sections as a function of the $\pT$ and the $|y|$ of both the top quark and the top
antiquark, measured at the parton level~\cite{CMS:2016xnv}. The
dominant systematic uncertainties are data statistics, \tch\
renormalisation and factorisation scales, top-quark mass variation, JES
and jet energy resolution (JER). All measurements for all
centre-of-mass energies are compared to various Monte Carlo (MC) predictions as
well as to fixed-order QCD calculations where available, and all are
in agreement with the SM prediction.

\subsection{Measurement of the \Wt\ production cross-section}

The latest results on the \Wt\ production cross-sections include the
combination of cross-section measurements at \cmenergyEight\ by the ATLAS and CMS
Collaborations~\cite{ATLAS:2016wds}. In this channel, events are
selected by requiring two opposite-sign high-$\pT$ isolated leptons (electrons or muons) and
one or two jets, where at least one must be a $b$-jet. The
two measurements used in the combination are based on
integrated luminosities of 20.3~fb$^{-1}$ and 12.2~fb$^{-1}$,
respectively. The results are combined using the best linear unbiased
estimator method and the cross-section is determined as $23.1 \pm
1.1~{\rm (stat.)} \pm 3.3~{\rm (syst.)}$~pb. The dominant systematic uncertainties are the
theory modelling and the jet uncertainties. At \cmenergyThirteen, the
early measurement of the inclusive \Wt\ production cross-section using
3.2~fb$^{-1}$ is presented by the ATLAS
Collaboration~\cite{Aaboud:2016lpj}. The \Wt\ signal is separated from the \ttbar\ background using boosted
decision tree (BDT) discriminants. The cross-section is
extracted by fitting templates to the data distributions, and is
measured to be $94 \pm 10~{\rm (stat.)} ^{+28}_{-22} {\rm
  (syst.)}$~pb. All measurements are in agreement with the NLO+NNLL expectation.

\subsection{Evidence of the \sch\ production}

The \sch\ signal is characterised by one high-$\pT$ isolated charged
lepton (electron or muon), large $\Met$ and two jets, where
both must be identified as $b$-jets. After the event
signal selection, the main backgrounds are \ttbar\ and $W$+jets
production. The most recent analyses are performed using integrated
luminosities of 5.1~fb$^{-1}$ at \cmenergySeven\ and 19.7~fb$^{-1}$ at
\cmenergyEight\ in the CMS analysis~\cite{Khachatryan:2016ewo} and with 20.3~fb$^{-1}$ at \cmenergyEight\ in the ATLAS
analysis~\cite{Aad:2015upn}. In order to separate the signal from the large
background contributions a profile maximum-likelihood fit is performed
to data, based on a given discriminant. In the CMS analysis a BDT
discriminant constructed in the signal region and in the \ttbar\
control region is used. In the ATLAS analysis, a combined discriminant based on
the matrix element method~\cite{Kondo:1988yd, Kondo:1991dw} and
on the lepton charge is used. The results presented by the CMS
Collaboration correspond to an observed (expected) significance of 0.9
(0.5) and 2.3 (0.8) standard deviations at \cmenergyCombRunI,
respectively. This allows to provide upper limits for the \sch\ production, which are 31.4~pb at
\cmenergySeven\ and 28.8~pb at \cmenergyEight, both at 95\%
confidence level (CL). The ATLAS Collaboration presented
the first evidence of the \sch\ single-top-quark production at the
LHC, measuring the cross section of $4.8 \pm 0.8~{\rm
  (stat.)}~{}^{+1.6}_{-1.3}~{\rm (syst.)}$~pb at \cmenergyEight, with
an observed (expected) significance of 3.2 (3.9) standard
deviations. Dominating uncertainties are \ttbar\ factorisation and
renormalisation scales and JES/JER in the CMS analysis and MC
statistics, JER, and the modelling of the \tch\ process in the ATLAS analysis.

\subsection{Summary of single-top-quark production cross-sections}

\begin{figure}[!htb]
  \centering
  \includegraphics[width=0.70\linewidth]{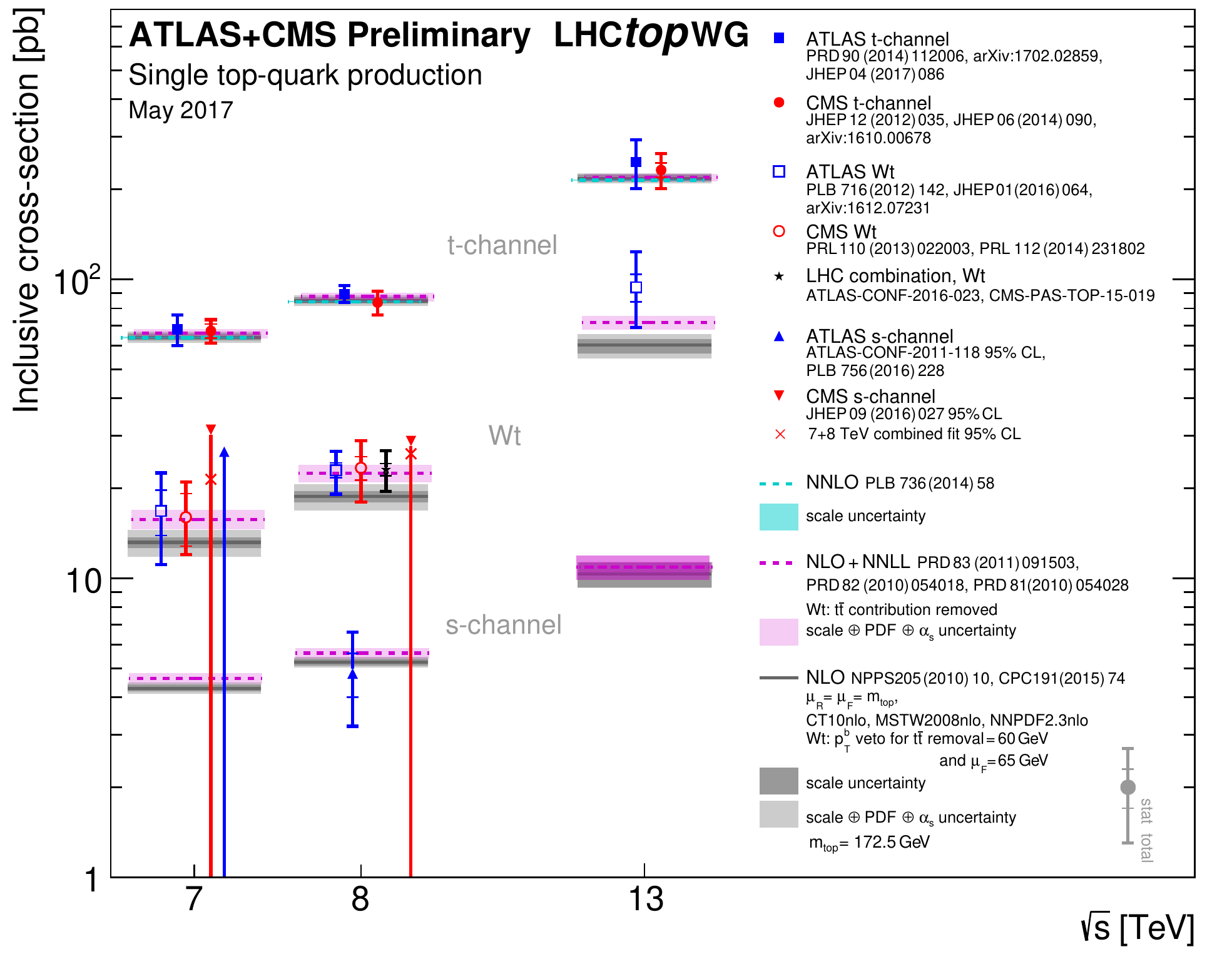}
  \caption{Summary of ATLAS and CMS measurements of the
    single-top-quark production cross-sections as
    a function of the centre-of-mass energy~\cite{LHCTopWGSummaryPlots}. The measurements are
    compared to theoretical calculations based on NLO, NLO+NNLL and NNLO (\tch\ only).}
  \label{fig:summary}
\end{figure}

\Figure{fig:summary} summarises the LHC single-top-quark cross-section
measurements at \cmenergyCombAll\ as a function of the centre-of-mass energy~\cite{LHCTopWGSummaryPlots}. For the CMS measurement of
the \sch\ only an upper limit is shown. The measurements are compared
to theoretical calculations based on NLO, NLO+NNLL and
next-to-next-to-leading order (NNLO).

\subsection{$\Vtb$ determination}

\begin{figure}[!htb]
  \centering
  \includegraphics[width=0.7\linewidth]{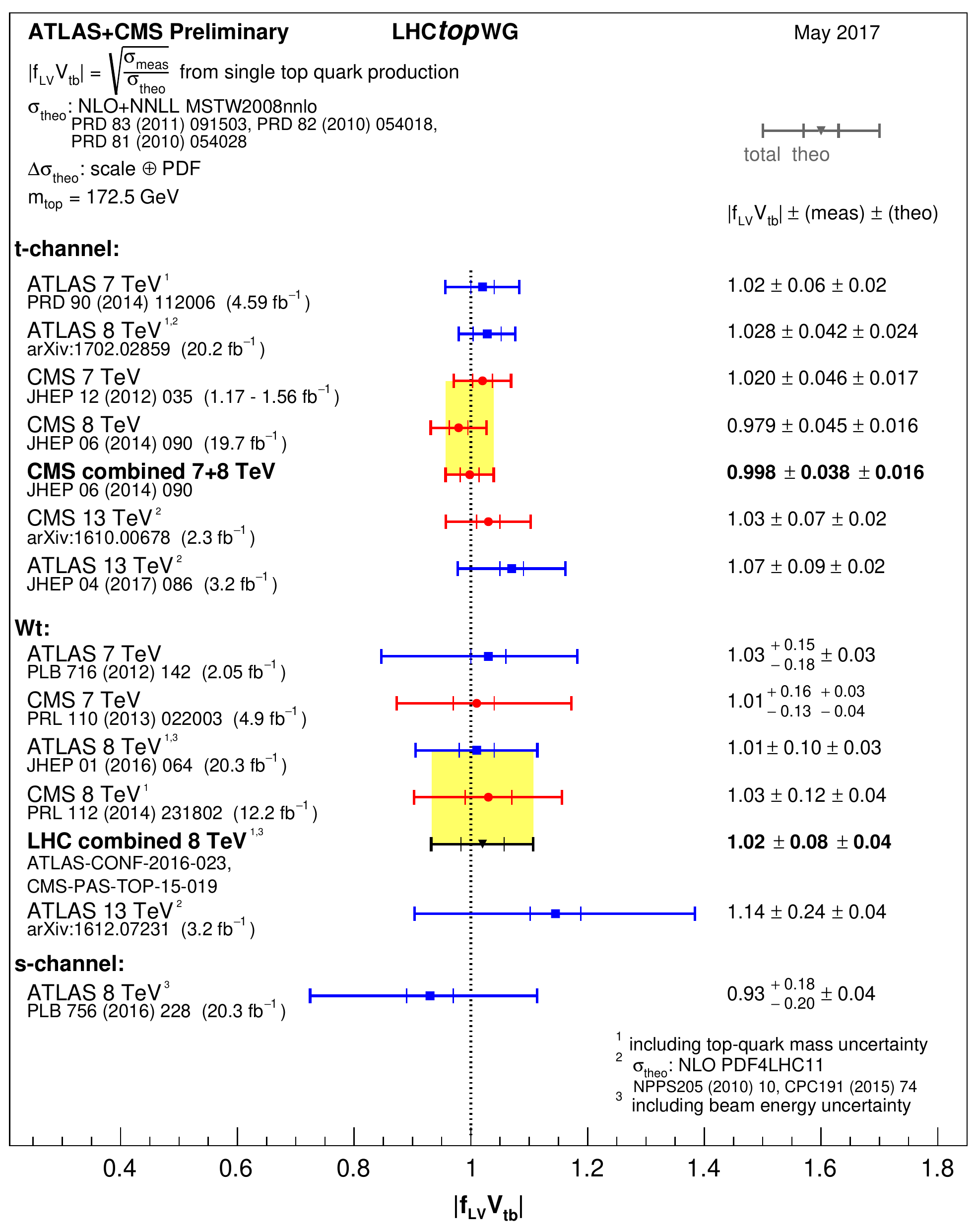}
  \caption{Summary of the ATLAS and CMS extractions of the CKM matrix
    element \flvVtb\ from single-top-quark measurements~\cite{LHCTopWGSummaryPlots}. Uncertainties
    originating from theoretical predictions and experimental measurement of the
    cross-sections are shown.
    % For each result, the contribution to the total uncertainty originating from
    % the uncertainty on the theoretical prediction for the
    % single-top-quark production cross-section is shown along with the
    % uncertainty originating from the experimental measurement of the
    % cross-section.
  }
  \label{fig:Vtb}
\end{figure}

Single-top-quark production provides a direct probe of the SM \Wtb\
coupling at the production vertex. In particular, a direct estimate of
$\Vtb$ can be obtained from the single-top-quark cross-section
measurement $\sigmameas$ together with its corresponding theoretical
expectation $\sigmatheo$, as $\flvVtb^2 = \sigmameas/\sigmatheo$,
%\begin{eqnarray}
%\label{eq:vtb}
%\Vtb = \sqrt{\frac{\sigmameas}{\sigmatheo}},
%\end{eqnarray}
where the $\flv$ term is a model-independent form factor for the
left-handed vector coupling. This factor is exactly one in the SM while it can change significantly in
the presence of new phenomena. The estimate only assumes that
$\Vts,\Vtd \ll \Vtb$~\cite{Alwall:2006bx}, and that the
\Wtb\ interaction involves a left-handed weak coupling like that in
the SM. However, no assumption is made about the number of quark
generations or unitarity of the CKM matrix, $\Vts^2+\Vtd^2+\Vtb^2=1$. In the
SM, $\Vtb$ is very close to one and it is considered equal to one in
theory calculations for single-top-quark cross-sections. \Figure{fig:Vtb} shows the summary of
the ATLAS and CMS extractions of the CKM matrix element \flvVtb\ from
single-top-quark measurements~\cite{LHCTopWGSummaryPlots}.

\subsection{Search for rare measurement of the \tZch\ production cross-section}

A search for the production of a single top quark in association with
a $Z$ boson, \tZch, using an integrated luminosity of 19.7~fb$^{-1}$
at \cmenergyEight\ is presented by the CMS
Collaboration~\cite{Sirunyan:2017kkr}. This analysis identifies the
expected SM process and searches for flavour-changing neutral-current (FCNC)
interactions. Final states with three leptons (electrons or muons),
large \Met\ and at least one jet are investigated. The signal is
extracted from data by performing a simultaneous binned
maximum-likelihood fit to the BDT discriminant distributions of the signal
samples and a background-enriched control region. Event yields
compatible with \tZch\ SM production are observed, and the
corresponding cross-section is measured to be $10^{+8}_{-7}$~fb, which
is agreement with the SM prediction of 8.2~fb~\cite{Campbell:2013yla}. Data
statistics is the dominant uncertainty. The observed
(expected) significance is 2.4 (1.8) standard deviations. For the FCNC
search, the SM \tZch\ process is considered as a background. No
evidence for \tZch-FCNC interactions is found, and limits at 95\% CL
are set on the branching fraction (BR) for the decay of a top quark
into a $Z$ boson and a quark. The observed (expected) limits are BR$(t \rightarrow Zu) <$~0.022\% (0.027\%) and BR$(t \rightarrow
Zc) <$~0.049\% (0.118\%), which improve the previous limits set by the
CMS Collaboration by about a factor of two.

\section{Measurement of the top-quark and $W$ boson spin observables}

\begin{figure}[!htbp]
  \centering
  \subfloat[]{ \includegraphics[width=0.45\linewidth]{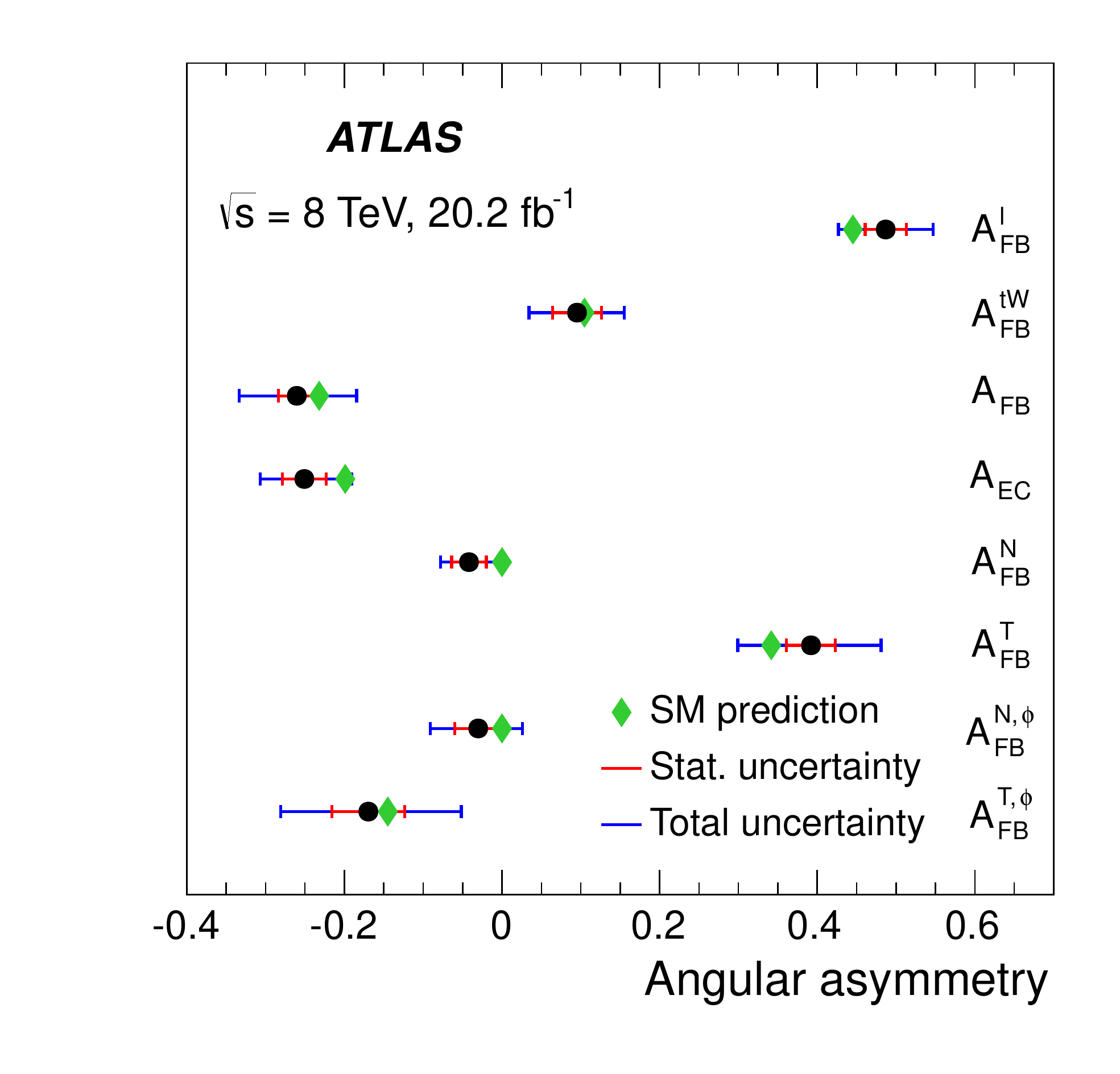}}
  \subfloat[]{ \includegraphics[width=0.45\linewidth]{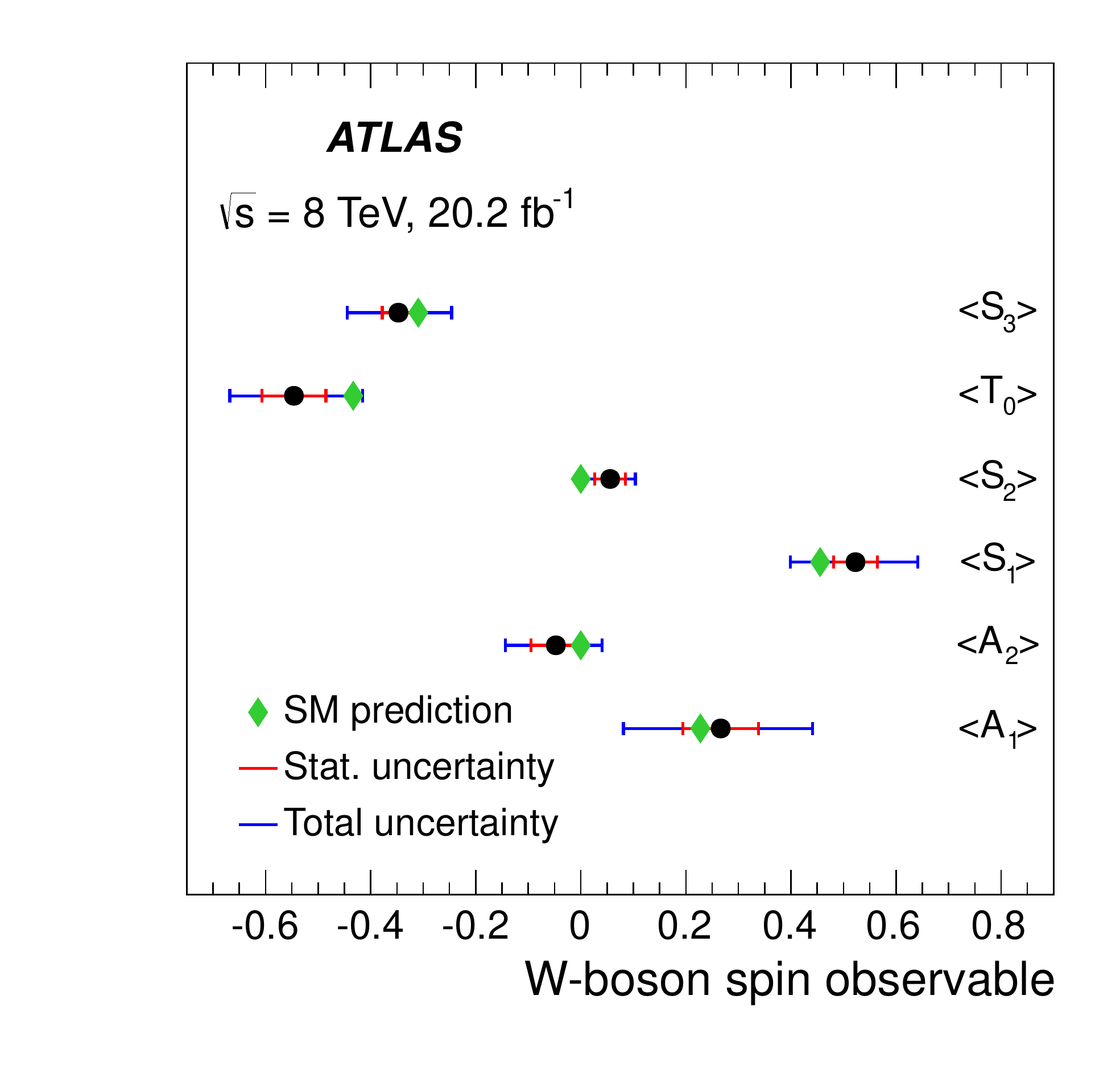} }
  \caption{Summary of (a) the measured asymmetries and (b) the
    measured values of the $W$ boson spin observables, both compared with the SM predictions~\cite{Aaboud:2017aqp}.}
  \label{fig:topPandWspin}
\end{figure}

Measurements of the top-quark and $W$ boson polarisation observables in \tch\ single-top-quark production
at \cmenergyEight\ with 20.2~fb$^{-1}$ are presented by the ATLAS
Collaboration~\cite{Aaboud:2017aqp}. The events containing the \tch\
signature are considered. A cut-based analysis is used
to discriminate the signal events from background. The polarisation
observables are measured from asymmetries in various angular
distributions unfolded to the parton level. Unfolding corrections based on a SM simulation of the \tch\ process
are used, as well as model-independent corrections derived through an
interpolation method. All the measured asymmetries and the measured $W$ boson spin observables are
shown in \Figure{fig:topPandWspin}. Additionally, limits on \imgr\
from a model-independent measurement are also set, giving $\imgr \in
[-0.18, 0.06]$ at the 95\% CL. This assumes $\vl = 1$ and that all anomalous couplings other than \imgr\ vanish. The \ttbar\ modelling,
JES and MC statistics are the dominant systematic uncertainties. Results are in agreement with SM
predictions~\cite{Aguilar-Saavedra:2015yza}.

\section{Analysis of the \Wtb\ vertex at production and decay}

\begin{figure}[!htbp]
  \centering
  \subfloat[]{ \includegraphics[width=0.52\linewidth]{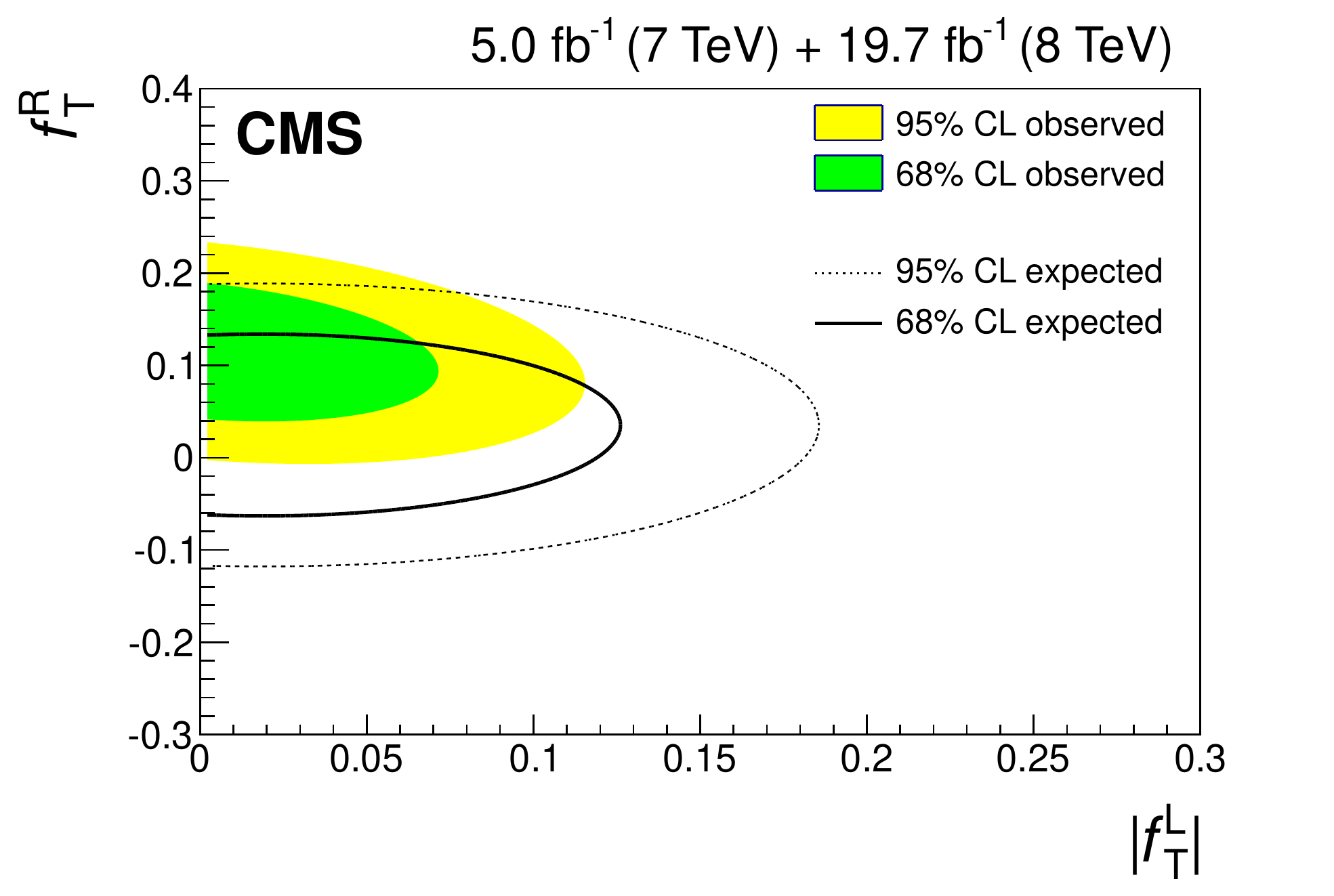}}
  %\subfloat[]{ \includegraphics[width=0.45\linewidth]{P.pdf}}
  \subfloat[]{ \includegraphics[width=0.45\linewidth]{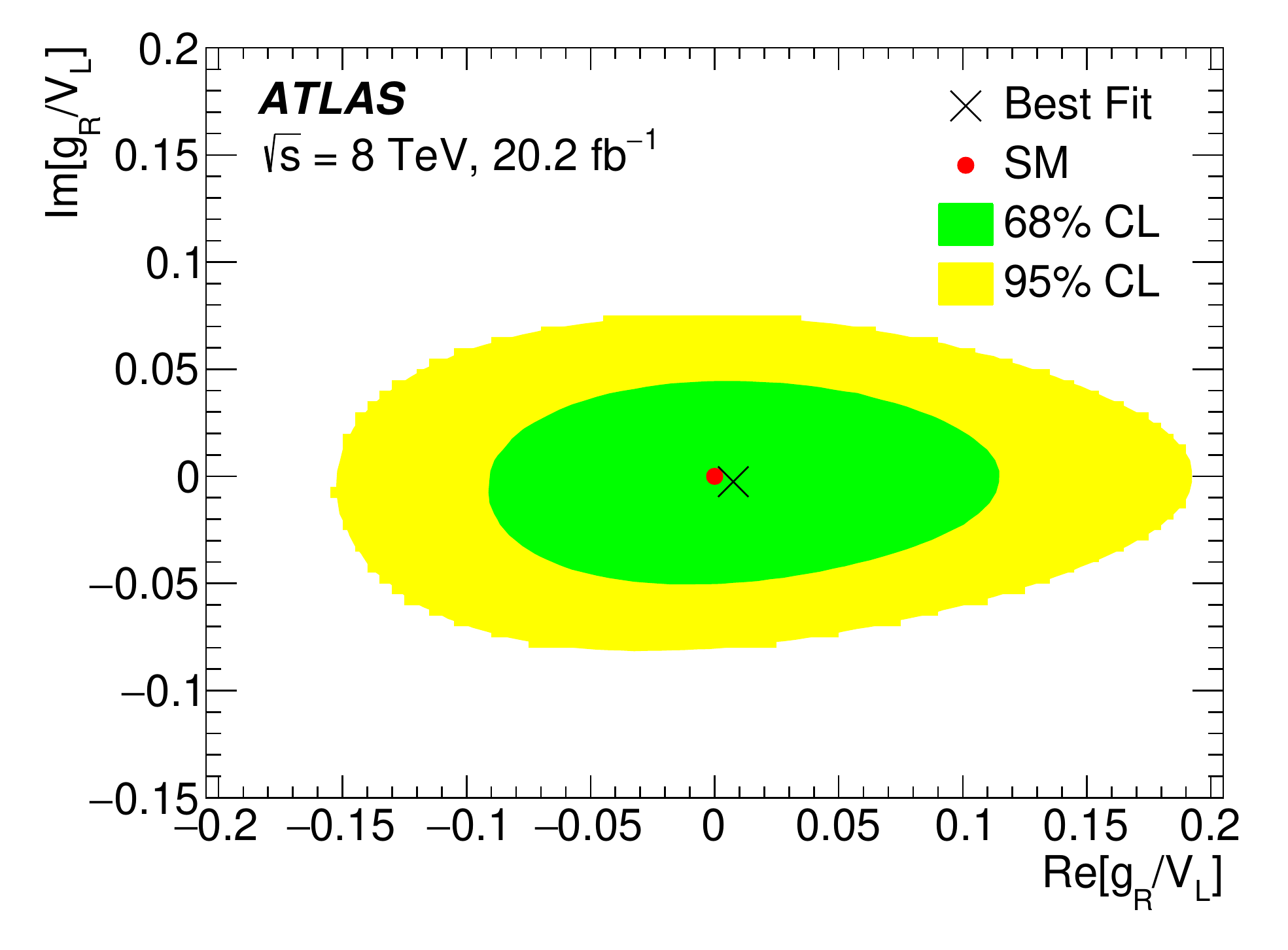}}
  % \caption{The likelihood profile for the top-quark polarisation (a)
  %   is shown. The black line indicates the evaluated likelihood in each bin of the profiled variable. The joint likelihood contour of $\Im{\gr/\vl}$ as a function
  %   of $\Re{\gr/\vl}$ (b) is shown. The black x mark indicates the observed
  %   value. The red point (a) or red dashed line (b) represent the SM
  %   expectation. In both cases, the 68\% and 95\% CL regions are shown in green and yellow, respectively~\cite{Aaboud:2017yqf}.}
  \caption{The result (a) of the three-dimensional simultaneous fit
    on the $\fvl \equiv \Re{\vl}$, $\ftl \equiv \Re{\gl}$ and $\ftr \equiv \Re{\gr}$ couplings in the two-dimension plane
    ($|\ftl| \equiv |\gl|$ and $\ftr \equiv \Re{\gr}$) is
    shown~\cite{Khachatryan:2016sib}. The solid and dotted lines correspond to the 68\% and 95\% CL expected regions,
    respectively. Additionally (b), the joint likelihood contour of $\Im{\gr/\vl}$ as a function
    of $\Re{\gr/\vl}$ (b) is shown~\cite{Aaboud:2017yqf}. The black x mark indicates the observed
    value and the red point represents the SM expectation. In
    both cases, the 68\% and 95\% CL observed regions are shown in green and
    yellow, respectively.}
  \label{fig:3A}
\end{figure}

Details studies of the  \Wtb\ vertex at production and decay are shown
in this section. The CMS Collaboration considers the events produced in the \tch\ to set limits
on anomalous \Wtb\ couplings and to search for top-quark FCNC
interactions~\cite{Khachatryan:2016sib}. Integrated luminosities of 5.0~fb$^{-1}$ at \cmenergySeven\ and 19.7~fb$^{-1}$ at
\cmenergyEight\ are used. A Bayesian NN technique is used to
discriminate between the signal and backgrounds in SM and different
non-SM scenarios. Simultaneous fits are performed on two or
three NN discriminants. The results of the fits, as shown in
\Figure{fig:3A}(a), are presented in the form of
two- or three-dimensional contours. The 95\% CL exclusion limits on
(real) anomalous couplings are measured to be
$|\vr| < 0.16$, $|\gl| < 0.057$, and $\gr \in [-0.049, 0.048]$. These are obtained by integrating over the other anomalous
parameter in the corresponding scenario, assuming $\vl = 1$ and that
the remaining anomalous couplings other than
ones being evaluated are set to their SM expectation. For the FCNC
search, the measured 95\% CL upper limits are $|\kappa_{tug}|/\Lambda < 4.1 × 10^{-3}$~TeV$^{-1}$ and
$|\kappa_{tcg}|/\Lambda < 1.8 \times 10^{-2}$~TeV$^{-1}$, where
$\kappa_{tug}$ and $\kappa_{tcg}$ define the strength of the FCNC
interactions in the $tug$ or $tcg$ vertices and $\Lambda$ is the scale
for new physics. These correspond to the upper
limits of BR$(t \rightarrow ug) = 2.0  \times 10^{-5}$ and BR$(t
\rightarrow cg) = 4.1 \times 10^{-4}$.

The ATLAS Collaboration also considers events containing the \tch\
signature to use an orthogonal series density estimation technique to
perform an angular analysis~\cite{Aaboud:2017yqf} of a triple-differential decay rate in \tch\ production to
simultaneously determine five generalised helicity fractions and
phases, as well as the top-quark polarisation. The
(complex) anomalous couplings are then constrained. This cut-based
analysis uses 20.2~fb$^{-1}$ at \cmenergyEight. Detector effects are
deconvolved from data using Fourier techniques and a multidimensional
likelihood function is obtained. From numerical calculations of this
likelihood function, measurements, limits and distributions as the one
shown in \Figure{fig:3A}(b) are obtained. The fraction of decays containing transversely polarised $W$ bosons is 
  measured to be $0.30 \pm 0.02~{\rm (stat.)} \pm 0.05~{\rm
    (syst.)}$. The phase between amplitudes for transversely and longitudinally
  polarised $W$ bosons recoiling against left-handed $b$-quarks is
  measured to be $0.002\pi \pm 0.0014\pi~{\rm (stat.)} \pm 0.011\pi~{\rm
    (syst.)}$, giving no indication of CP violation.  The fractions of longitudinal or transverse $W$ bosons accompanied by right-handed 
  $b$-quarks are also constrained. Based on these measurements, $\Re{\gr/\vl} \in [-0.12, 0.17]$ and
  $\Im{\gr/\vl} \in [-0.07, 0.06]$ at 95\% CL. Constraints are also
  placed on the other couplings; $|\vr/\vl| < 0.37$ and $|\gl/\vl| < 0.29$ at 95\% CL. In addition, the top-quark
  polarisation is constrained to be $P > 0.72$ at 95\% CL. None of the
  above ATLAS measurements make assumptions about the value of any of the 
  other parameters or couplings. In both cases, ATLAS and CMS results are in agreement with the SM.

\section{Conclusions}

The ATLAS and CMS Collaborations have produced high-precision
cross-section measurements at \cmenergyCombAll\ using collision data
from the LHC, including fiducial and detailed differential studies. A search for rare \tZch\ production is also
presented. Additionally, the top-quark polarisation and $W$ spin
observables are measured and complete analyses of the \Wtb\ vertex at
production and decay are presented. All these measurements are in good
agreement with predictions and no hint for physics beyond the SM is
observed so far in the top-quark sector.

%%%%%%%%%%%%%%%%%%%%%%%%%%%%%%%%%%%%%%%%%%%%%%%%%%%%%%%%%%%%%%%%%%%%%%%%%%%

%%  if necessary
\Acknowledgements
The author would like to thank the ATLAS and
CMS Collaborations for their incredible work in single-top-quark
studies and also to the organisers for the kind invitation to this
extraordinary conference.

%%%%%%%%%%%%%%%%%%%%%%%%%%%%%%%%%%%%%%%%%%%%%%%%%%%%%%%%%%%%%%%%%%%%%%%%%%%


\begin{thebibliography}{99}

%%
%%  bibliographic items can be constructed using the LaTeX format in SPIRES:
%%    see    http://www.slac.stanford.edu/spires/hep/latex.html
%%  SPIRES will also supply the CITATION line information; please include it.
%%

%\cite{Aaltonen:2009jj}
\bibitem{Aaltonen:2009jj} 
  %%T.~Aaltonen {\it et al.} [CDF Collaboration],
  CDF Collaboration,
  %``First Observation of Electroweak Single Top Quark Production,''
  Phys.\ Rev.\ Lett.\  {\bf 103}, 092002 (2009)
  %%doi:10.1103/PhysRevLett.103.092002
  [arXiv:0903.0885 [hep-ex]].
  %%CITATION = doi:10.1103/PhysRevLett.103.092002;%%
  %444 citations counted in INSPIRE as of 28 Aug 2017

%\cite{Abazov:2009ii}
\bibitem{Abazov:2009ii} 
  % V.~M.~Abazov {\it et al.} [D0 Collaboration],
  D0 Collaboration,
  %``Observation of Single Top Quark Production,''
  Phys.\ Rev.\ Lett.\  {\bf 103}, 092001 (2009)
  %%doi:10.1103/PhysRevLett.103.092001
  [arXiv:0903.0850 [hep-ex]].
  %%CITATION = doi:10.1103/PhysRevLett.103.092001;%%
  %438 citations counted in INSPIRE as of 28 Aug 2017

%\cite{ATLAS:2014wva}
\bibitem{ATLAS:2014wva} 
  ATLAS and CDF and CMS and D0 Collaborations,
  %``First combination of Tevatron and LHC measurements of the top-quark mass,''
  arXiv:1403.4427 [hep-ex].
  %%CITATION = ARXIV:1403.4427;%%
  %475 citations counted in INSPIRE as of 28 Aug 2017

%\cite{Bigi:1986jk}
\bibitem{Bigi:1986jk} 
  I.~I.~Y.~Bigi, Y.~L.~Dokshitzer, V.~A.~Khoze, J.~H.~Kuhn and P.~M.~Zerwas,
  %``Production and Decay Properties of Ultraheavy Quarks,''
  Phys.\ Lett.\ B {\bf 181}, 157 (1986).
  %%doi:10.1016/0370-2693(86)91275-X
  %%CITATION = doi:10.1016/0370-2693(86)91275-X;%%
  %551 citations counted in INSPIRE as of 30 Aug 2017

%\cite{Aliev:2010zk}
\bibitem{Aliev:2010zk} 
  M.~Aliev, H.~Lacker, U.~Langenfeld, S.~Moch, P.~Uwer and M.~Wiedermann,
  %``HATHOR: HAdronic Top and Heavy quarks crOss section calculatoR,''
  Comput.\ Phys.\ Commun.\  {\bf 182}, 1034 (2011)
  %%doi:10.1016/j.cpc.2010.12.040
  [arXiv:1007.1327 [hep-ph]].
  %%CITATION = doi:10.1016/j.cpc.2010.12.040;%%
  %699 citations counted in INSPIRE as of 29 Aug 2017

%\cite{Kant:2014oha}
\bibitem{Kant:2014oha} 
  P.~Kant, O.~M.~Kind, T.~Kintscher, T.~Lohse, T.~Martini, S.~Mölbitz, P.~Rieck and P.~Uwer,
  %``HatHor  for single top-quark production: Updated predictions and uncertainty estimates for single top-quark production in hadronic collisions,''
  Comput.\ Phys.\ Commun.\  {\bf 191}, 74 (2015)
  %%doi:10.1016/j.cpc.2015.02.001
  [arXiv:1406.4403 [hep-ph]].
  %%CITATION = doi:10.1016/j.cpc.2015.02.001;%%
  %103 citations counted in INSPIRE as of 29 Aug 2017

%\cite{Kidonakis:2010ux}
\bibitem{Kidonakis:2010ux} 
  N.~Kidonakis,
  %``Two-loop soft anomalous dimensions for single top quark associated production with a W- or H-,''
  Phys.\ Rev.\ D {\bf 82}, 054018 (2010)
  %%doi:10.1103/PhysRevD.82.054018
  [arXiv:1005.4451 [hep-ph]].
  %%CITATION = doi:10.1103/PhysRevD.82.054018;%%
  %622 citations counted in INSPIRE as of 29 Aug 2017

%\cite{Kidonakis:2012rm}
\bibitem{Kidonakis:2012rm} 
  N.~Kidonakis,
  %``NNLL threshold resummation for top-pair and single-top production,''
  Phys.\ Part.\ Nucl.\  {\bf 45}, no. 4, 714 (2014)
  %%doi:10.1134/S1063779614040091
  [arXiv:1210.7813 [hep-ph]].
  %%CITATION = doi:10.1134/S1063779614040091;%%
  %111 citations counted in INSPIRE as of 29 Aug 2017

%\cite{Kidonakis:2013zqa}
\bibitem{Kidonakis:2013zqa} 
  N.~Kidonakis,
  %``Top Quark Production,''
  %%doi:10.3204/DESY-PROC-2013-03/Kidonakis
  arXiv:1311.0283 [hep-ph].
  %%CITATION = doi:10.3204/DESY-PROC-2013-03/Kidonakis;%%
  %61 citations counted in INSPIRE as of 29 Aug 2017

%\cite{Cabibbo:1963yz}
\bibitem{Cabibbo:1963yz} 
  N.~Cabibbo,
  %``Unitary Symmetry and Leptonic Decays,''
  Phys.\ Rev.\ Lett.\  {\bf 10}, 531 (1963).
  %%doi:10.1103/PhysRevLett.10.531
  %%CITATION = doi:10.1103/PhysRevLett.10.531;%%
  %5566 citations counted in INSPIRE as of 29 Aug 2017

%\cite{Kobayashi:1973fv}
\bibitem{Kobayashi:1973fv} 
  M.~Kobayashi and T.~Maskawa,
  %``CP Violation in the Renormalizable Theory of Weak Interaction,''
  Prog.\ Theor.\ Phys.\  {\bf 49}, 652 (1973).
  %%doi:10.1143/PTP.49.652
  %%CITATION = doi:10.1143/PTP.49.652;%%
  %9173 citations counted in INSPIRE as of 29 Aug 2017

%\cite{AguilarSaavedra:2008zc}
\bibitem{AguilarSaavedra:2008zc} 
  J.~A.~Aguilar-Saavedra,
  %``A Minimal set of top anomalous couplings,''
  Nucl.\ Phys.\ B {\bf 812}, 181 (2009)
  %%doi:10.1016/j.nuclphysb.2008.12.012
  [arXiv:0811.3842 [hep-ph]].
  %%CITATION = doi:10.1016/j.nuclphysb.2008.12.012;%%
  %256 citations counted in INSPIRE as of 30 Aug 2017

%\cite{Mahlon:1996pn}
\bibitem{Mahlon:1996pn} 
  G.~Mahlon and S.~J.~Parke,
  %``Improved spin basis for angular correlation studies in single top quark production at the Tevatron,''
  Phys.\ Rev.\ D {\bf 55}, 7249 (1997)
  %%doi:10.1103/PhysRevD.55.7249
  [hep-ph/9611367].
  %%CITATION = doi:10.1103/PhysRevD.55.7249;%%
  %163 citations counted in INSPIRE as of 30 Aug 2017

%\cite{Aguilar-Saavedra:2015yza}
\bibitem{Aguilar-Saavedra:2015yza} 
  J.~A.~Aguilar-Saavedra and J.~Bernabeu,
  %``Breaking down the entire W boson spin observables from its decay,''
  Phys.\ Rev.\ D {\bf 93}, no. 1, 011301 (2016)
  %%doi:10.1103/PhysRevD.93.011301
  [arXiv:1508.04592 [hep-ph]].
  %%CITATION = doi:10.1103/PhysRevD.93.011301;%%
  %15 citations counted in INSPIRE as of 30 Aug 2017

%\cite{Aad:2008zzm}
\bibitem{Aad:2008zzm} 
  %%G.~Aad {\it et al.} [ATLAS Collaboration],
  ATLAS Collaboration,
  %``The ATLAS Experiment at the CERN Large Hadron Collider,''
  JINST {\bf 3}, S08003 (2008).
  %%doi:10.1088/1748-0221/3/08/S08003
  %%CITATION = doi:10.1088/1748-0221/3/08/S08003;%%
  %5659 citations counted in INSPIRE as of 28 Aug 2017

%\cite{Chatrchyan:2008aa}
\bibitem{Chatrchyan:2008aa} 
  %%S.~Chatrchyan {\it et al.} [CMS Collaboration],
  CMS Collaboration,
  %``The CMS Experiment at the CERN LHC,''
  JINST {\bf 3}, S08004 (2008).
  %%doi:10.1088/1748-0221/3/08/S08004
  %%CITATION = doi:10.1088/1748-0221/3/08/S08004;%%
  %4468 citations counted in INSPIRE as of 28 Aug 2017

%%%%%%%%%%%%%%%%%%%%%%%%%%%%%%%%%%%%%%%%%%%%%%%%%%%%%%%%%%%%%%%%%%%%%%%%%%%

%\cite{CMS:2015jca}
\bibitem{CMS:2015jca} 
  CMS Collaboration,
  %``Fiducial t channel single top-quark cross section at 8 TeV,''
  CMS-PAS-TOP-15-007.
  %%CITATION = CMS-PAS-TOP-15-007;%%
  %7 citations counted in INSPIRE as of 29 Aug 2017

%\cite{Aaboud:2017pdi}
\bibitem{Aaboud:2017pdi} 
  %%M.~Aaboud {\it et al.} [ATLAS Collaboration],
  ATLAS Collaboration,
  %``Fiducial, total and differential cross-section measurements of $t$-channel single top-quark production in $pp$ collisions at 8 TeV using data collected by the ATLAS detector,''
  Eur.\ Phys.\ J.\ C {\bf 77}, no. 8, 531 (2017)
  %%doi:10.1140/epjc/s10052-017-5061-9
  [arXiv:1702.02859 [hep-ex]].
  %%CITATION = doi:10.1140/epjc/s10052-017-5061-9;%%
  %4 citations counted in INSPIRE as of 29 Aug 2017

%\cite{Sirunyan:2016cdg}
\bibitem{Sirunyan:2016cdg} 
  %%A.~M.~Sirunyan {\it et al.} [CMS Collaboration],
  CMS Collaboration,
  %``Cross section measurement of $t$-channel single top quark production in pp collisions at $\sqrt s =$ 13 TeV,''
  Phys.\ Lett.\ B {\bf 772}, 752 (2017)
  %%doi:10.1016/j.physletb.2017.07.047
  [arXiv:1610.00678 [hep-ex]].
  %%CITATION = doi:10.1016/j.physletb.2017.07.047;%%
  %19 citations counted in INSPIRE as of 29 Aug 2017

%\cite{Aaboud:2016ymp}
\bibitem{Aaboud:2016ymp} 
  %%M.~Aaboud {\it et al.} [ATLAS Collaboration],
  ATLAS Collaboration,
  %``Measurement of the inclusive cross-sections of single top-quark and top-antiquark $t$-channel production in $pp$ collisions at $\sqrt{s}$ = 13 TeV with the ATLAS detector,''
  JHEP {\bf 1704}, 086 (2017)
  %%doi:10.1007/JHEP04(2017)086
  [arXiv:1609.03920 [hep-ex]].
  %%CITATION = doi:10.1007/JHEP04(2017)086;%%
  %16 citations counted in INSPIRE as of 29 Aug 2017

%\cite{CMS:2016xnv}
\bibitem{CMS:2016xnv} 
  CMS Collaboration,
  %``Measurement of the differential cross section for $t$-channel single-top-quark production at $\sqrt{s}=13~\mathrm{TeV}$,''
  CMS-PAS-TOP-16-004.
  %%CITATION = CMS-PAS-TOP-16-004;%%
  %10 citations counted in INSPIRE as of 29 Aug 2017

%%%%%%%%%%%%%%%%%%%%%%%%%%%%%%%%%%%%%%%%%%%%%%%%%%%%%%%%%%%%%%%%%%%%%%%%%%%

%\cite{ATLAS:2016wds}
\bibitem{ATLAS:2016wds} 
  ATLAS collaboration,
  %``Combination of cross-section measurements for associated production of a single top-quark and a W boson at √s = 8 TeV with the ATLAS and CMS experiments,''
  ATLAS-CONF-2016-023.
  %%CITATION = ATLAS-CONF-2016-023;%%
  %3 citations counted in INSPIRE as of 29 Aug 2017

%\cite{Aaboud:2016lpj}
\bibitem{Aaboud:2016lpj} 
  %%M.~Aaboud {\it et al.} [ATLAS Collaboration],
  ATLAS Collaboration,
  %``Measurement of the cross-section for producing a $W$ boson in association with a single top quark in $pp$ collisions at $\sqrt{s}={13} \text{TeV}$ with ATLAS,''
  arXiv:1612.07231 [hep-ex].
  %%CITATION = ARXIV:1612.07231;%%
  %5 citations counted in INSPIRE as of 29 Aug 2017

%%%%%%%%%%%%%%%%%%%%%%%%%%%%%%%%%%%%%%%%%%%%%%%%%%%%%%%%%%%%%%%%%%%%%%%%%%%

%\cite{Khachatryan:2016ewo}
\bibitem{Khachatryan:2016ewo} 
  %%V.~Khachatryan {\it et al.} [CMS Collaboration],
  CMS Collaboration,
  %``Search for s channel single top quark production in pp collisions at $ \sqrt{s}=7 $ and 8 TeV,''
  JHEP {\bf 1609}, 027 (2016)
  %%doi:10.1007/JHEP09(2016)027
  [arXiv:1603.02555 [hep-ex]].
  %%CITATION = doi:10.1007/JHEP09(2016)027;%%
  %23 citations counted in INSPIRE as of 29 Aug 2017

%\cite{Aad:2015upn}
\bibitem{Aad:2015upn} 
  %%G.~Aad {\it et al.} [ATLAS Collaboration],
  ATLAS Collaboration,
  %``Evidence for single top-quark production in the $s$-channel in proton-proton collisions at $\sqrt{s}=$8 TeV with the ATLAS detector using the Matrix Element Method,''
  Phys.\ Lett.\ B {\bf 756}, 228 (2016)
  %%doi:10.1016/j.physletb.2016.03.017
  [arXiv:1511.05980 [hep-ex]].
  %%CITATION = doi:10.1016/j.physletb.2016.03.017;%%
  %27 citations counted in INSPIRE as of 29 Aug 2017

%\cite{Kondo:1988yd}
\bibitem{Kondo:1988yd} 
  K.~Kondo,
  %``Dynamical Likelihood Method for Reconstruction of Events With Missing Momentum. 1: Method and Toy Models,''
  J.\ Phys.\ Soc.\ Jap.\  {\bf 57}, 4126 (1988).
  %%doi:10.1143/JPSJ.57.4126
  %%CITATION = doi:10.1143/JPSJ.57.4126;%%
  %152 citations counted in INSPIRE as of 29 Aug 2017

%\cite{Kondo:1991dw}
\bibitem{Kondo:1991dw} 
  K.~Kondo,
  %``Dynamical likelihood method for reconstruction of events with missing momentum. 2: Mass spectra for 2 ---> 2 processes,''
  J.\ Phys.\ Soc.\ Jap.\  {\bf 60}, 836 (1991).
  %%doi:10.1143/JPSJ.60.836
  %%CITATION = doi:10.1143/JPSJ.60.836;%%
  %96 citations counted in INSPIRE as of 29 Aug 2017

%%%%%%%%%%%%%%%%%%%%%%%%%%%%%%%%%%%%%%%%%%%%%%%%%%%%%%%%%%%%%%%%%%%%%%%%%%%

%\cite{LHCTopWGSummaryPlots}
\bibitem{LHCTopWGSummaryPlots} 
 LHC$top$WG Collaboration,
  %``LHCTopWG Summary Plots''
  https://twiki.cern.ch/twiki/bin/view/LHCPhysics/LHCTopWGSummaryPlots.

%%%%%%%%%%%%%%%%%%%%%%%%%%%%%%%%%%%%%%%%%%%%%%%%%%%%%%%%%%%%%%%%%%%%%%%%%%%

%\cite{Alwall:2006bx}
\bibitem{Alwall:2006bx} 
  J.~Alwall {\it et al.},
  %``Is V($_{tb}$) $\simeq$ 1?,''
  Eur.\ Phys.\ J.\ C {\bf 49}, 791 (2007)
  %%doi:10.1140/epjc/s10052-006-0137-y
  [hep-ph/0607115].
  %%CITATION = doi:10.1140/epjc/s10052-006-0137-y;%%
  %187 citations counted in INSPIRE as of 29 Aug 2017

%%%%%%%%%%%%%%%%%%%%%%%%%%%%%%%%%%%%%%%%%%%%%%%%%%%%%%%%%%%%%%%%%%%%%%%%%%%

%\cite{Sirunyan:2017kkr}
\bibitem{Sirunyan:2017kkr} 
  %%A.~M.~Sirunyan {\it et al.} [CMS Collaboration],
  CMS Collaboration,
  %``Search for associated production of a Z boson with a single top quark and for tZ flavour-changing interactions in pp collisions at $ \sqrt{s}=8 $ TeV,''
  JHEP {\bf 1707}, 003 (2017)
  %%doi:10.1007/JHEP07(2017)003
  [arXiv:1702.01404 [hep-ex]].
  %%CITATION = doi:10.1007/JHEP07(2017)003;%%
  %6 citations counted in INSPIRE as of 29 Aug 2017

%\cite{Campbell:2013yla}
\bibitem{Campbell:2013yla} 
  J.~Campbell, R.~K.~Ellis and R.~Röntsch,
  %``Single top production in association with a Z boson at the LHC,''
  Phys.\ Rev.\ D {\bf 87}, 114006 (2013)
  %%doi:10.1103/PhysRevD.87.114006
  [arXiv:1302.3856 [hep-ph]].
  %%CITATION = doi:10.1103/PhysRevD.87.114006;%%
  %54 citations counted in INSPIRE as of 29 Aug 2017

%%%%%%%%%%%%%%%%%%%%%%%%%%%%%%%%%%%%%%%%%%%%%%%%%%%%%%%%%%%%%%%%%%%%%%%%%%%

%\cite{Aaboud:2017aqp}
\bibitem{Aaboud:2017aqp} 
  %%M.~Aaboud {\it et al.} [ATLAS Collaboration],
  ATLAS Collaboration, 
  %``Probing the W tb vertex structure in t-channel single-top-quark production and decay in pp collisions at $ \sqrt{s}=8 $ TeV with the ATLAS detector,''
  JHEP {\bf 04}, 124 (2017)
  %%doi:10.1007/JHEP04(2017)124
  [arXiv:1702.08309 [hep-ex]].
  %%CITATION = doi:10.1007/JHEP04(2017)124;%%
  %5 citations counted in INSPIRE as of 29 Aug 2017

%%%%%%%%%%%%%%%%%%%%%%%%%%%%%%%%%%%%%%%%%%%%%%%%%%%%%%%%%%%%%%%%%%%%%%%%%%%

%\cite{Khachatryan:2016sib}
\bibitem{Khachatryan:2016sib} 
  %%V.~Khachatryan {\it et al.} [CMS Collaboration],
  CMS Collaboration,
  %``Search for anomalous Wtb couplings and flavour-changing neutral currents in t-channel single top quark production in pp collisions at $\sqrt{s} =$ 7 and 8 TeV,''
  JHEP {\bf 1702}, 028 (2017)
  %%doi:10.1007/JHEP02(2017)028
  [arXiv:1610.03545 [hep-ex]].
  %%CITATION = doi:10.1007/JHEP02(2017)028;%%
  %10 citations counted in INSPIRE as of 30 Aug 2017

%\cite{Aaboud:2017yqf}
\bibitem{Aaboud:2017yqf} 
  %%M.~Aaboud {\it et al.} [ATLAS Collaboration],
  ATLAS Collaboration,
  %``Analysis of the $Wtb$ vertex from the measurement of triple-differential angular decay rates of single top quarks produced in the $t$-channel at $\sqrt{s}$ = 8 TeV with the ATLAS detector,''
  arXiv:1707.05393 [hep-ex].
  %%CITATION = ARXIV:1707.05393;%%

%%%%%%%%%%%%%%%%%%%%%%%%%%%%%%%%%%%%%%%%%%%%%%%%%%%%%%%%%%%%%%%%%%%%%%%%%%%


\end{thebibliography}
\end{document}